\documentclass[aps,prb,twocolumn,groupedaddress]{revtex4}

\usepackage{graphicx}
\usepackage{afterpage}
\begin{document}

\title{Manganese valence and magnetotransport in ultrathin films of La$_{0.67}$Ca$_{0.33}$MnO$_3$}

\author{C. Beekman and J. Aarts.}
\affiliation{Kamerlingh Onnes Laboratory, Leiden University, P.O. Box 9504, 2300 RA
Leiden, the Netherlands}
\author{M. Porcu and H. Zandbergen}
\affiliation{National Centre for High Resolution Microscopy, Kavli Institute for
Nanoscience, Delft Technical University, Lorentzweg 1, 2628 CJ Delft, The
Netherlands}

\bibliographystyle{apsrev}
\date{\today}

\begin{abstract}\noindent
We report a comparative study of the properties of very thin films of
La$_{0.67}$Ca$_{0.33}$MnO$_3$ grown epitaxially under strain on flat SrTiO$_3$
(STO), lattice matched on NdGaO$_3$ (NGO), and strained on $1^{\circ}$-miscut STO.
We use transmission electron microscopy and electron energy loss spectroscopy to
study the microstructure, composition, and Mn valence state. Near the interface we
find no significant segregation, but a charge compensation layer where the valence
is enhanced over the nominal value of 3.3+, and a relaxation to this value over
several nanometer. The transport properties show well-known behavior for the films
on flat STO and NGO, namely values of the metal-insulator transition temperature
which are strongly (STO) or only little (NGO) reduced with respect to the bulk
value. The reduction in films on miscut STO however is less strong than in films on
flat STO, even though they appear similar as to strain state and interface layer.
Magnetically, we find reduced values of the saturation magnetization for the
strained films with respect to the bulk value, which cannot only be ascribed to the
interface layer.
\end{abstract}
 \maketitle

\section{Introduction}
Transition metal oxides with the perovskite structure are strongly correlated
electron systems which show diversity in physical properties caused by the
competition between charge, spin and orbital degrees of freedom. Much work has been
done in the last decade to understand the rich underlying physics of the correlated
and semi-localized 3d electrons and their interaction with the lattice. In
particular for the manganites, showing the colossal magnetoresistance effect (CMR)
connected to the combined metal-insulator (MI) and paramagnetic-to-ferromagnetic
transition, much progress has been made, as can be found in a number of reviews
\cite{tokura06,dagotto08,salamon01}. Still, even in bulk materials the picture is
still being refined, as was for instance shown recently in the discovery of the
existence of glassy correlated phases in single crystals of optimally doped
La$_{0.7}$Ca$_{0.3}$MnO$_3$ \cite{lynn07}. Also for thin films, a large amount of
work has gone into basic questions on their physical properties, with the
possibility of strain engineering as an issue of special interest. In bulk
manganites of the type (RE$_{1-x})$A$_x$MnO$_3$ (RE is a 3+ Rare Earth ion, A is a
2+ alkaline ion) and at fixed RE to A ratio and therefore the Mn$^{3+}$ to Mn$^{4+}$
ratio, the properties can be changed by varying the radius of the 2+ ion. This is
basically because the ion radius influences the structure of the coupled network of
MnO$_6$ octahedra, which changes the balance between the itinerancy of the Mn 3$d$
electrons, and the strength of the Jahn-Teller distortions which tend to trap
electrons on the Mn sites. In films this effect can be amplified by growing on a
substrate with a different lattice parameter, thereby putting the film under tensile
or compressive strain. For instance in films of La$_{0.7}$Ca$_{0.3}$MnO$_3$ (LCMO;
pseudocubic lattice parameter $a_c$ = 0.386~nm) on SrTiO$_3$ (STO; $a_c$ = 0.391~nm)
the temperature of the metal-insulator transition T$_{MI}$ goes down to 110~K for
films with a thickness of around 10~nm, compared to a bulk value of 260~K
\cite{aarts98,bibes01,yang04}. Below about 5~nm the MI transition rather abruptly
disappears, mainly because the large strain leads to different crystal structures in
the film \cite{yang04}. At the same time, magnetic measurements indicate the
presence of a dead layer in the LCMO/STO system of a few nm and nanoscale phase
separation \cite{bibes01}, and also in general a lowered value of the saturation
magnetization for larger thicknesses \cite{aarts98}.

Gaining more understanding about these ultrathin films in the regime around 10~nm is
of obvious interest. In principle they are suitable as gated devices, since
SrTiO$_3$ is a highly polarizable dielectric, but then more needs to be known about
the electronic structure of the film close to the interface. This is of particular
interest since the Mn-valency may be influenced by discontinuities in the charge
variation, as is now under extensive investigation for the case of SrTiO$_3$ /
LaAlO$_3$ \cite{ohtomo04}. Another question is the role of disorder and strain
relaxation. Here we present a comparative study of LCMO films grown on flat
SrTiO$_3$ (STO; tensile strain), NdGaO$_3$ (NGO; strain-free) substrates as well as
of on STO substrates which have an intentional misorientation of the surface normal
of 1$^{\circ}$ towards the [010] direction, which leads to a stepped structure with
a terrace length of about 25 nm. We investigate the microstructure, the atomic
composition, and the Mn-valence state using High Resolution Transmission Electron
Microscopy (HR-TEM), and Electron Energy Loss Spectroscopy (EELS), but also the
magnetic and magnetotransport properties of these same films. We find that basically
all films show the bulk $Pnma$ structure. For films on both flat and miscut STO, the
Mn-valence is slightly higher at the interface (about 0.1~electron charge), and
drops to the bulk value within 3 to 5~nm. For films on NGO we find the same, which
suggests that it is neither a different crystal structure, nor strain, but rather
the change in charge distribution when going from substrate to film which is driving
this variation. We then show from transport data that these film show the same
trends in $T_{MI}$ as found before, with a strong drop for films on flat STO and a
small variation for films on NGO. We also find that $T_{MI}$ for films on miscut STO
stays significantly higher. Finally we come back to the issue of magnetization and
show that the saturation values for films in this thickness regime are lower than
can be expected on the basis of the non-bulk interface layer alone.
\section{Experimental}
\begin{figure}[t]
\includegraphics[width=8.50cm,height=8.25cm,angle=0]{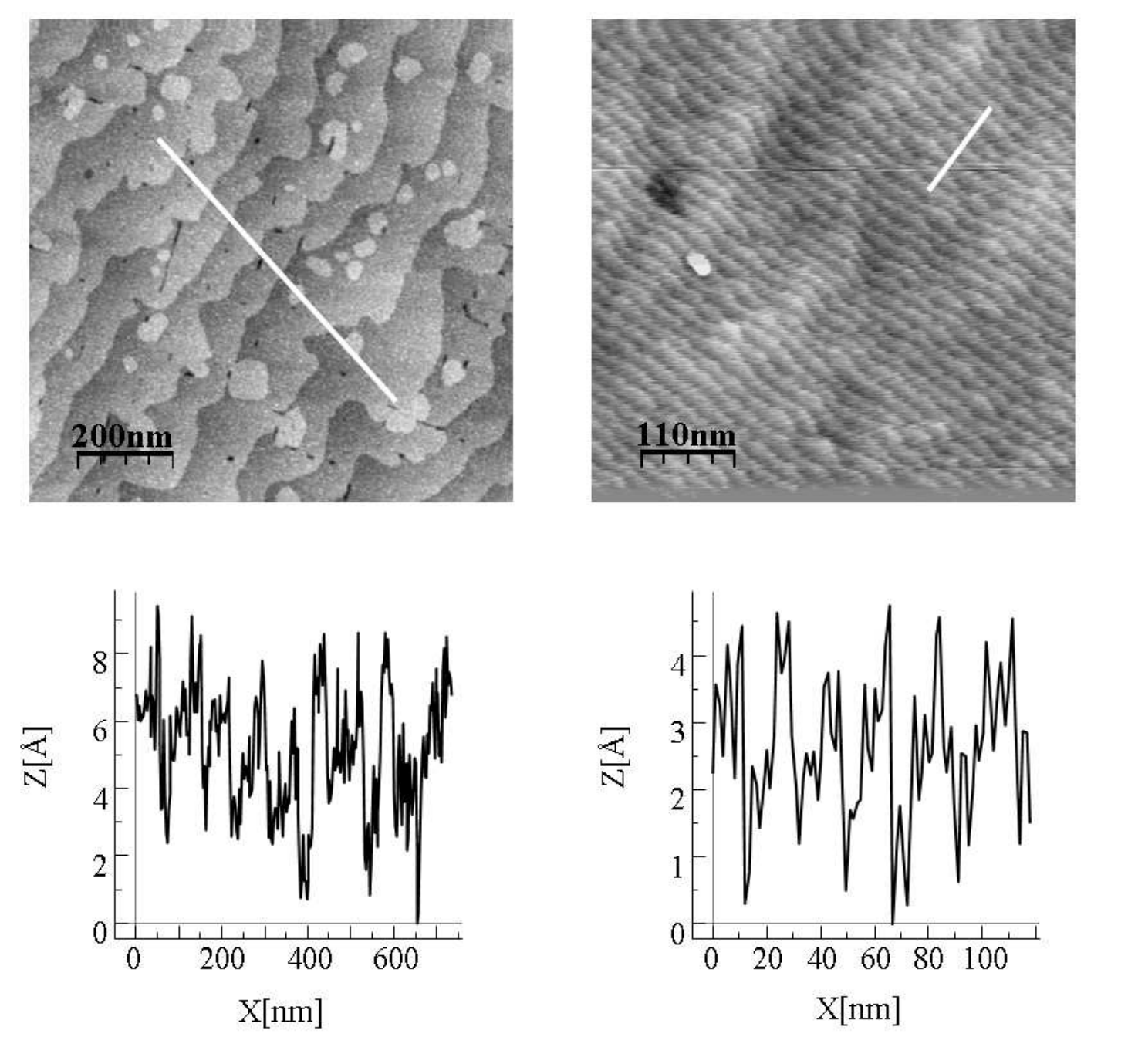}
\centering\caption{AFM images of LCMO films and corresponding profiles of (left) 15
nm LCMO on flat STO, L(15) (scale bar: 200 nm), and (right) 7 nm LCMO on STO with a
misorientation of 1$^{\circ}$ towards the [010] direction, L(7)$_{mis}$ (scale bar:
100 nm). }\label{fig1}
\end{figure}

\noindent
Epitaxial films of LCMO with thicknesses between 47 nm  and 6 nm were grown on (001)
STO substrates, by DC sputtering in oxygen pressure of 300 Pa at a growth
temperature of 840 $^{\circ}$C. The substrates have a misorientation of either
$<$0.2$^{\circ}$ in random direction which we denote as flat STO or 1$^{\circ}$
towards [010] direction. Here we define the nomenclature which we will use
throughout this paper to refer to our films. Films grown on flat STO are indicated
by L($d$), with $d$ the film thickness (rounded to the nearest integer value), films
grown on misoriented STO by L($d$)$_{mis}$ and films on NGO, by L($d$)$_{NGO}$.
Before and after film growth we used Atomic Force Microscopy (AFM) to asses the
quality of the STO substrate and the LCMO film (see Fig.\ref{fig1}).
One issue to be discussed is the surface termination of the STO, which can be either
a TiO$_2$- or an SrO-layer. Commercial substrates have mixed termination but can be
treated to become singly terminated, with the TiO$_2$ surface easier to fabricate
and more stable. Whether this is of influence on the film properties that are
investigated in this paper will be discussed later. All films show clear unit-cell
high step edges. The films grown on flat STO show an average terrace length of ~75
nm and the films grown on misoriented STO show an average terrace length of 20 nm,
identical to the terrace length of the substrate (see Fig. \ref{fig1}). The
thickness, growth rate and lattice parameters of the LCMO films were determined by
x-ray reflectivity (XRR) and reciprocal space mapping (RSM) measurements,
respectively. The average growth rate of our LCMO thin films is 0.8 nm/min, which
results in films with roughness of the order of the dimensions of the unit cell.

We have characterized the microstructure, the atomic composition, and the Mn-valence
state of LCMO thin films using the HR-TEM (High Resolution Transmission Electron
Microscopy) and EELS (Electron Energy Loss Spectroscopy) techniques. We will show
results of LCMO films grown on STO (flat), STO (1$^{\circ}$) and NGO. HR-TEM
specimens were prepared according to a standard cross-section preparation method.
Before insertion into the microscope, the specimens were plasma-cleaned for 1 minute
to prevent carbon contamination during the experiments. The analysis was performed
with a FEI TITAN equipped with a spherical aberration (C$s$) corrector and a High
Resolution Gatan Image Filter (HR-GIF) operated at 300kV. EELS data were collected
in scanning TEM mode with a probe size of about 0.2 - 0.5 nm. The spectra were
acquired by probing the same region only once to reduce the beam-induced damage. The
energy dispersion was 0.1 eV/channel for the Zero Loss Peak and 0.2 eV/channel for
the Mn L-edge to obtain more signal.

The magnetotransport properties were studied by measuring the current ($I$) -
voltage ($V$) characteristics as function of temperature and in high magnetic
fields. We used a PPMS (Physical Properties Measurement System, Quantum Design) for
temperature control (T = 20 - 300 K) and for magnetic field control (H$_a$ = 0 - 9
T), in combination with an external current source and nanovoltmeter. For the
magnetization measurements we used an MPMS (Magnetic Properties Measurement System,
Quantum Design) with T = 10 - 300 K and H$_a$ = 0 - 5T.

\section{Microstructure and Mn valency}\label{HR-TEM}
\subsection{Microstructure}\noindent
From HR-TEM investigations on several specimens for each film we find that our films
are epitaxial. The perovskite crystal structure of the films is close to cubic with
lattice parameter a$_c$ = 0.39 nm, but due to small rotations of the oxygen
octahedra it becomes orthorhombic (space group: $Pnma$). Using electron diffraction
we observed that throughout the films the bulk $Pnma$ structure is present with
lattice parameters of $\sqrt{2}a_c$, 2$a_c$ and $\sqrt{2}a_c$.
\begin{figure}[h!]
\begin{tabular}{c}
\includegraphics[width=8cm,height=2.7cm,angle=0]{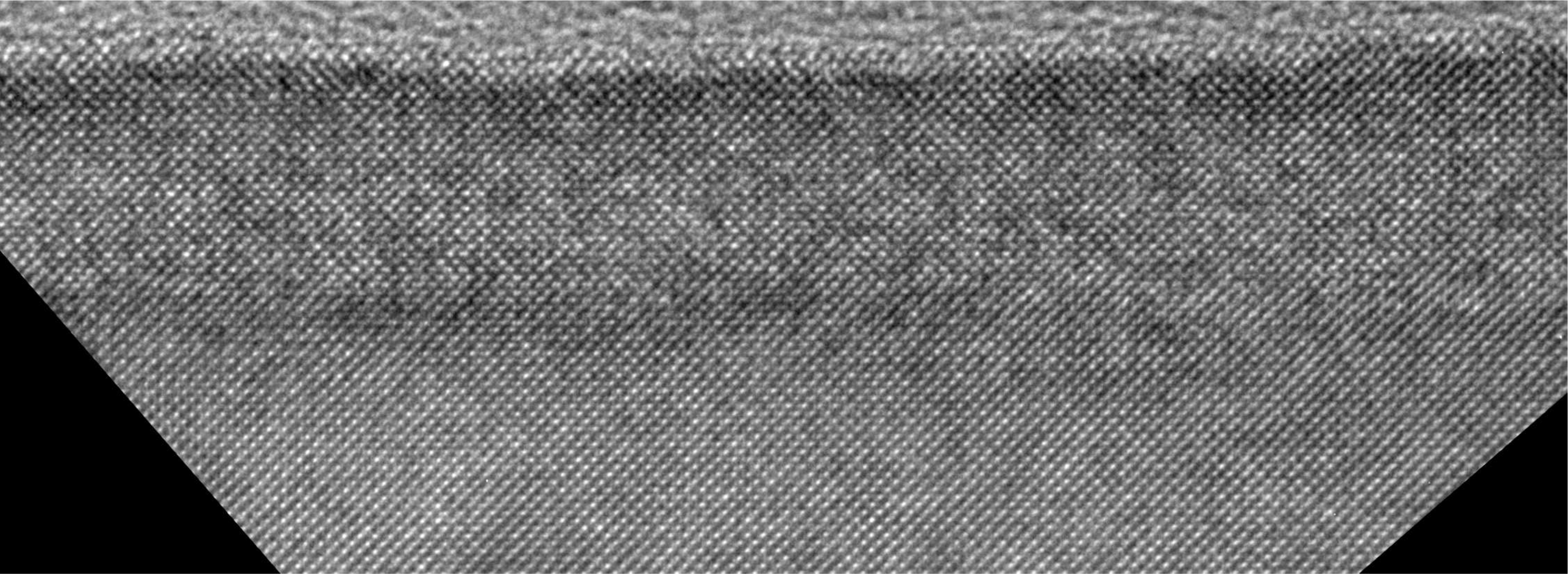} \\
\includegraphics[width=8cm,height=2.7cm,angle=0]{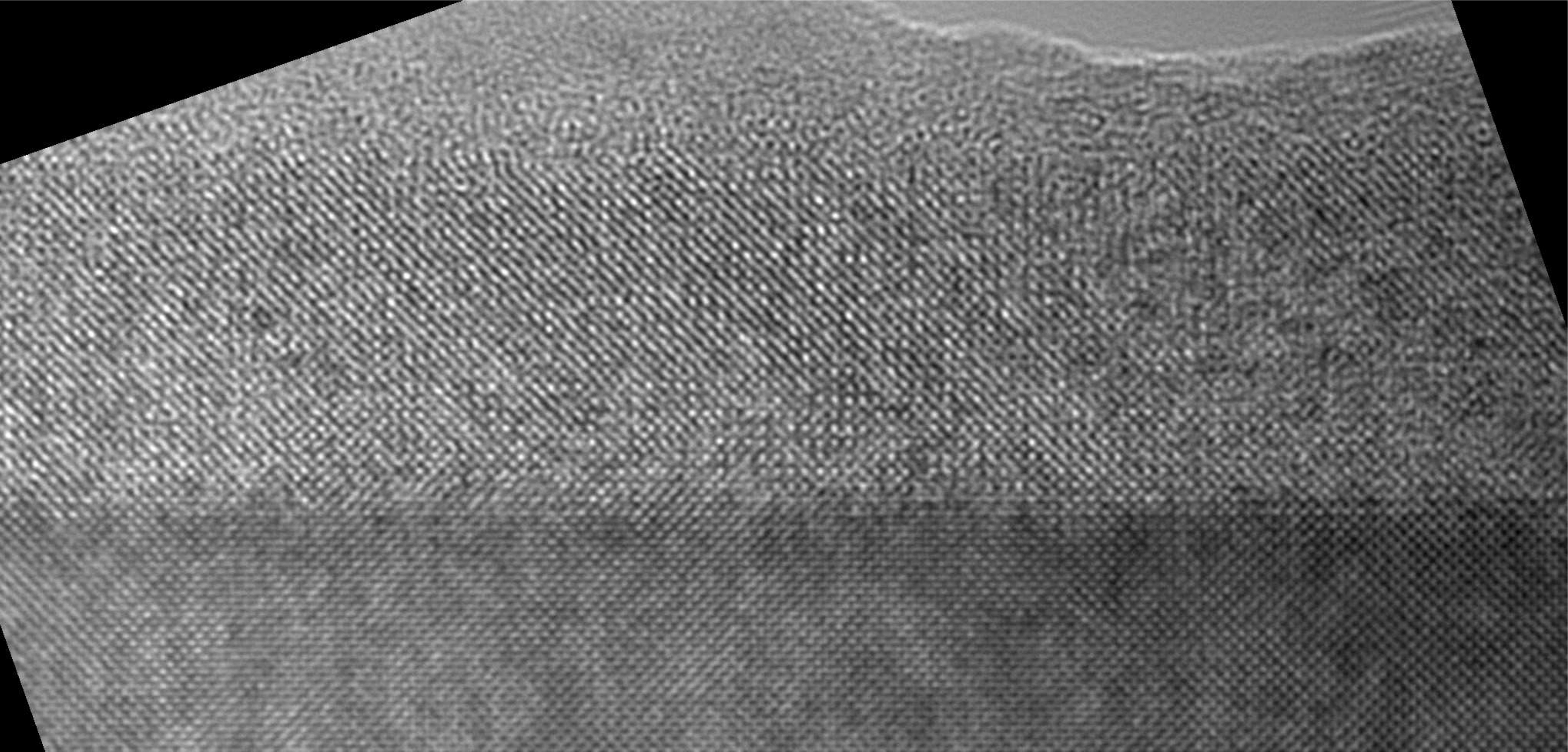} \\
\includegraphics[width=8cm,height=2.7cm,angle=0]{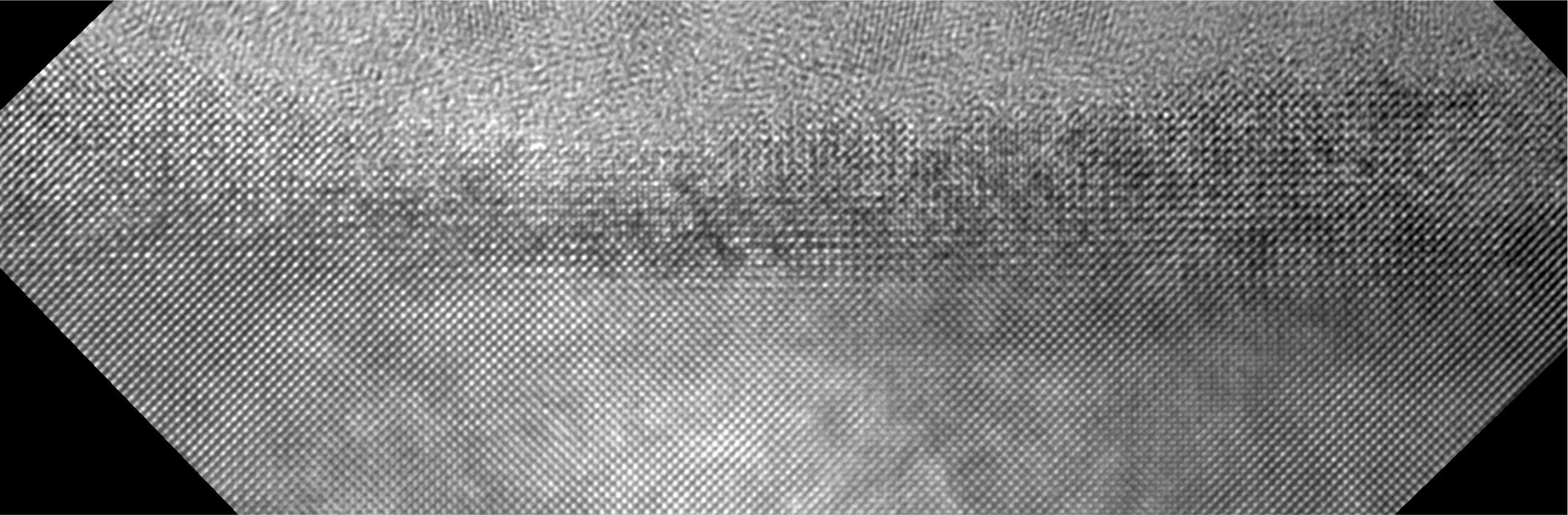} \\
\end{tabular}
\centering\caption{HR-TEM micrographs of LCMO films on STO, from top to bottom L(6),
L(10), L(7)$_{mis}$. The atom columns which are clearly visible in all cases set the
scale.}\label{fig2}
\end{figure}
For most films, the $b$ axis was found to be parallel to the interface normal (with
length 2$a_c$ in pseudocubic notation \cite{lat}). In HR-TEM mode (see
Fig.\ref{fig2}), and scanning along the interface, we did not observe any antiphase
boundaries or any domain type disorder, which is in line with previous reports
\cite{yang3a}. The amorphous layer visible at the top surface of the thin film in
Fig.\ref{fig2} (top image) is glue used during preparation of the sample. The HR-TEM
images shown in Fig.\ref{fig2} (samples L(6), L(10) and L(7)) all show a fully
epitaxial thin film. There is a special reason to show a micrograph for sample
L(10), since this film has deviating properties compared to the typical film on flat
STO, as will be shown later. HREM on this film showed no differences compared to the
other films. We also observe that films grown on 1$^{\circ}$ misoriented STO
(Fig.\ref{fig2}c) are epitaxial with no clear influence of the step edges on the
microstructure of the LCMO film. We also do not observe any misfit dislocation at
the film-substrate interface in any of the films, which were investigated with
HR-TEM.
\begin{figure}[t]
\centering\includegraphics[width=9.28cm,height=8cm,angle=0]{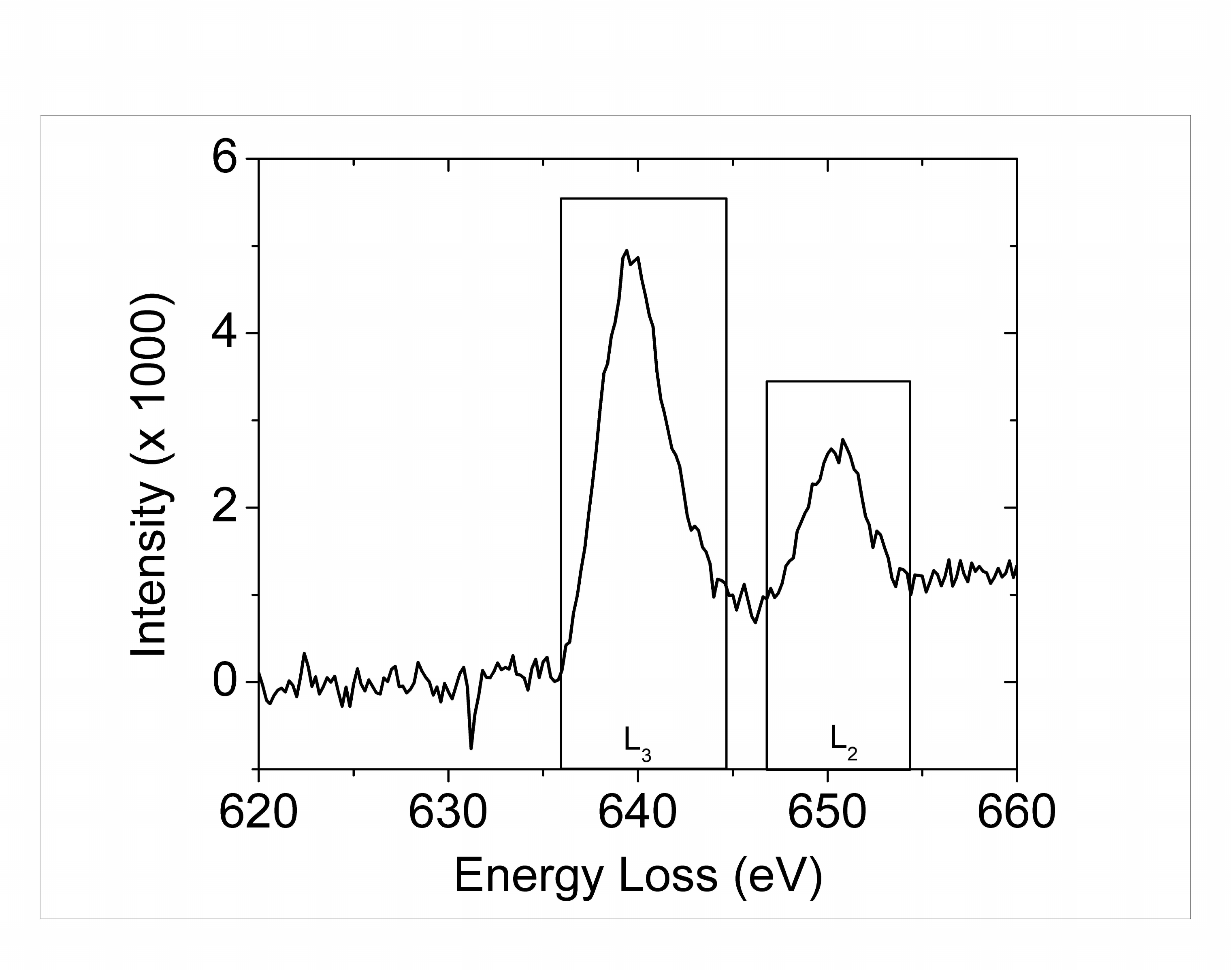}
\centering\caption{An EELS spectrum showing the electron intensity versus energy
loss, with the Mn L$_2$ and L$_3$ peaks indicated. The L$_{32}$ ratio is determined
from the integrated intensities ratio of the two peaks.}\label{fig3}
\end{figure}
\subsection{Mn-oxidation state and elemental
composition}\label{EELS}\noindent We have used EELS to investigate the composition
and the Mn-oxidation state across the thickness of our thin films. In an EELS
measurement the sample is exposed to an electron beam with a well defined (small
range of) kinetic energies. While the electrons go through the sample, inelastic
interactions (i.e. atom core loss, inner shell ionization) result in the loss of
kinetic energy, which can be measured using an electron spectrometer. An EELS
spectrum consists of a Zero Loss peak (ZLP), and subsequent peaks at lower energy
corresponding to different losses due to interaction with the sample. An example of
an EELS spectrum at the Mn L-edge is given in Fig.\ref{fig3}. The L$_2$ and L$_3$
peaks are indicated, and used to determine the Mn valence. The raw EELS data were
corrected for specimen thickness, which influences the ratio between intensities of
ZLP peak and the rest of the spectrum \cite{thickness3}. From the data it can be
assumed that the elemental composition will not have any dependence on the specimen
thickness. Since the only interest is to qualitatively monitor the changes across
the film, the cross section was taken as a constant. By acquiring the ZLP, Mn L-edge
and the La K-edge at 100~eV and calculating the ratio between the various integrals,
the Mn-valence and the elemental concentration across the film thickness are
obtained. The L$_{32}$ ratio, correlated to the Mn-valence, is determined from the
integrated intensities ratio L$_3$/L$_2$. Ratios of 2.1 and 2.7 correspond to
Mn$^{4+}$ and Mn$^{3+}$, respectively; the calculated valence for
La$_{0.7}$Ca$_{0.3}$MnO$_3$ is 3.3+ for bulk specimens \cite{Gianluigi3}. Acquiring
spectra  across the film yields elemental and valence profiles.
\begin{figure}[t]
\centering\includegraphics[width=8.9cm]{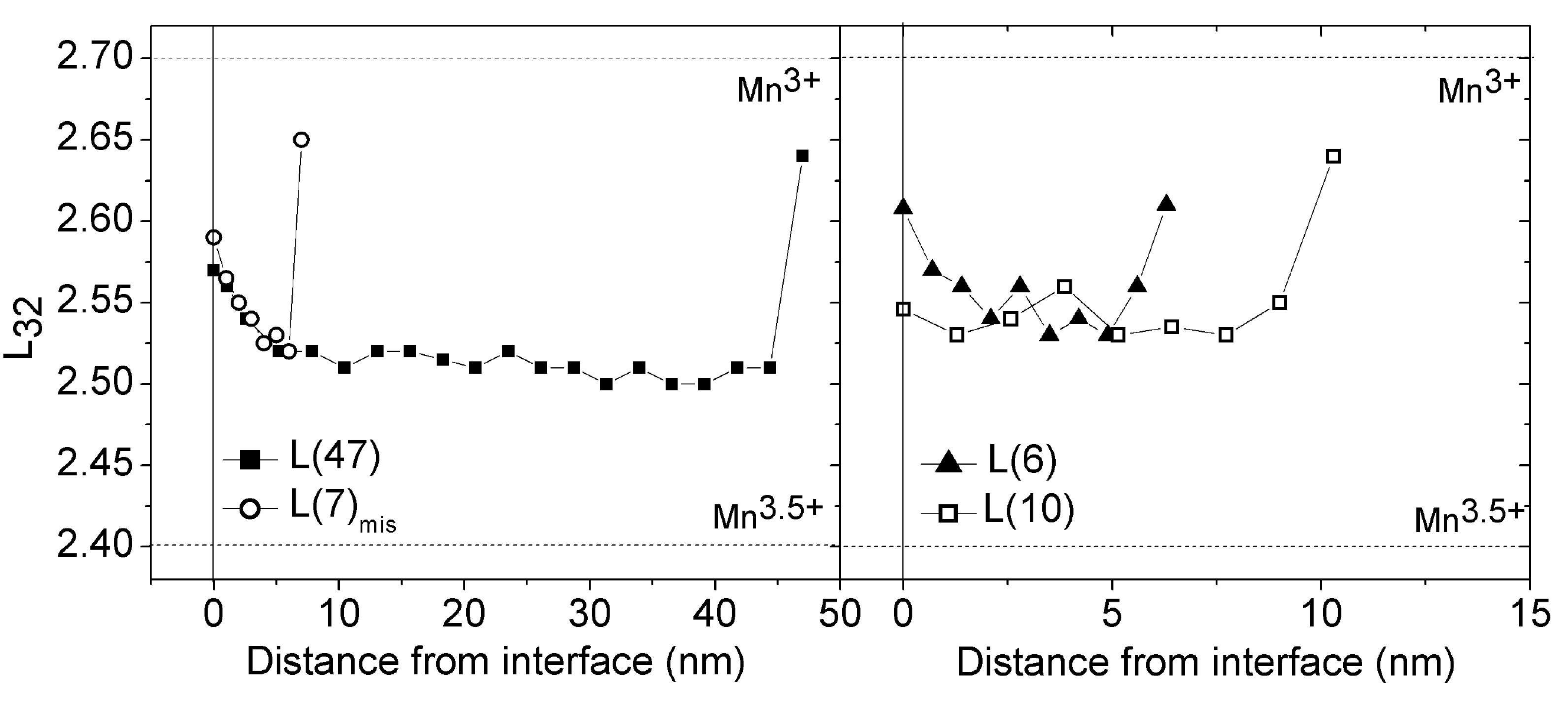}
\centering\caption{The calculated L$_{32}$ ratio from the EELS spectra. Left:
results for samples L(47) on flat STO and L(7)$_{mis}$ on miscut STO. Right: results
for L(6) and L(10) on flat STO. The drawn lines are guides to the eye. The dashed
lines indicate the L$_{32}$ values which correspond to Mn$^{3+}$ and Mn$^{3.5+}$
respectively. The solid line indicates the substrate-film interface. }\label{fig4}
\end{figure}

\begin{figure}[t]
\begin{tabular}{cc}
\includegraphics[width=4cm]{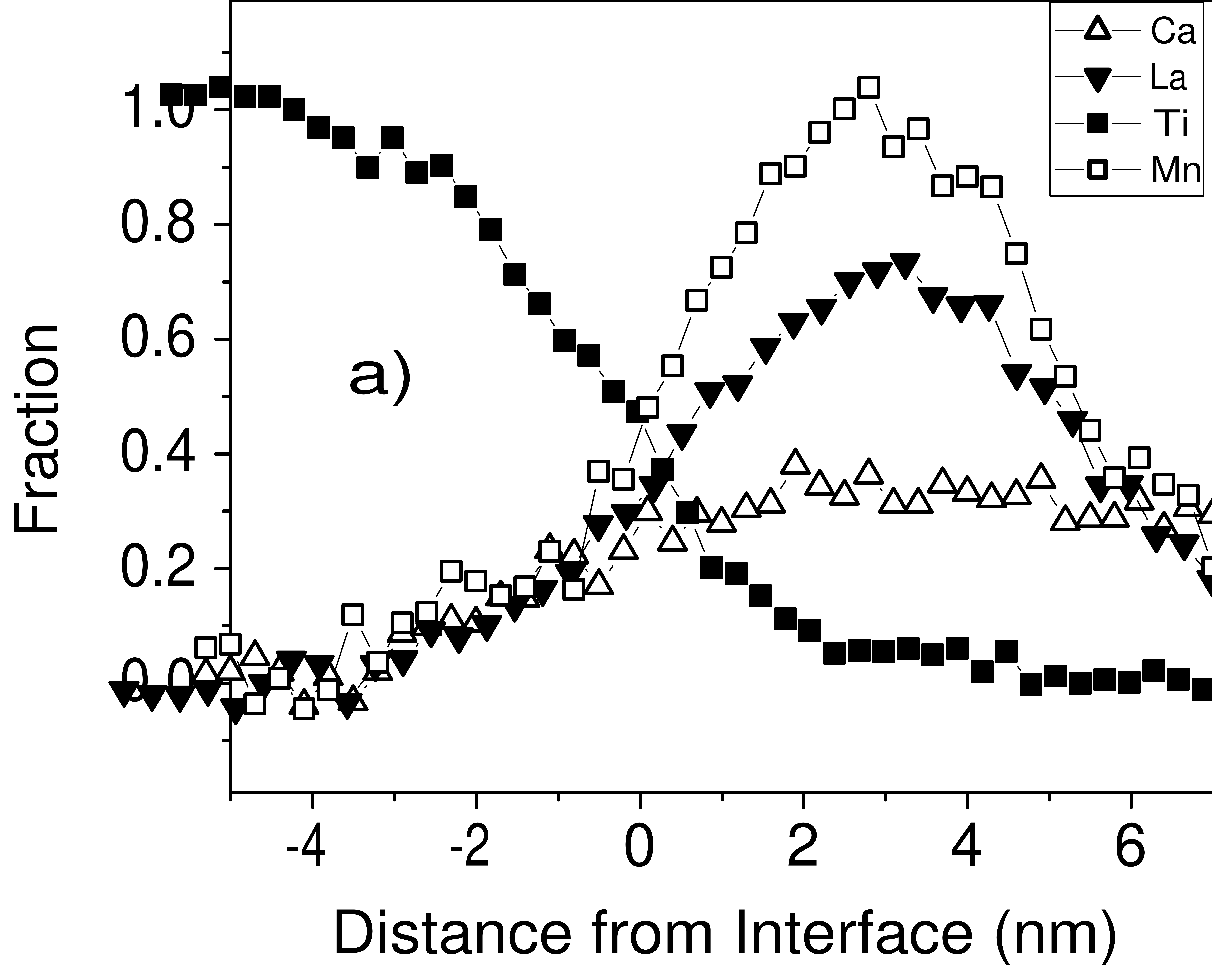} &
\includegraphics[width=4cm]{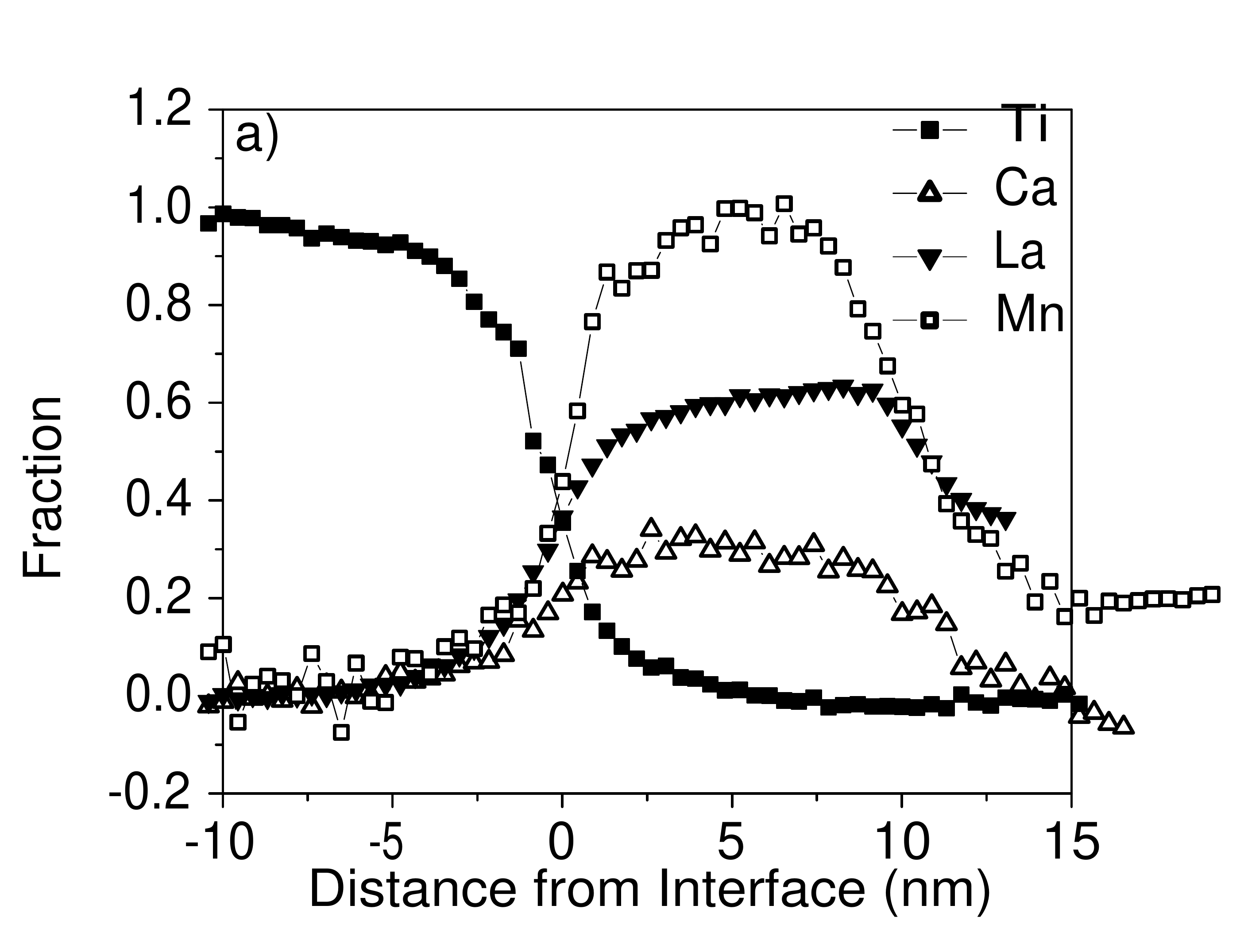} \\
\end{tabular}
\centering\caption{Typical elemental composition of a) sample L(6) and b) sample
L(10), plotted as elemental fraction versus distance from the interface. The
elements shown are Ca (open triangles), La (filled triangles), Ti (filled squares),
Mn (open squares) }\label{fig5}
\end{figure}

\begin{figure}[t]
\centering\includegraphics[width=9cm,height=3.98cm,angle=0]{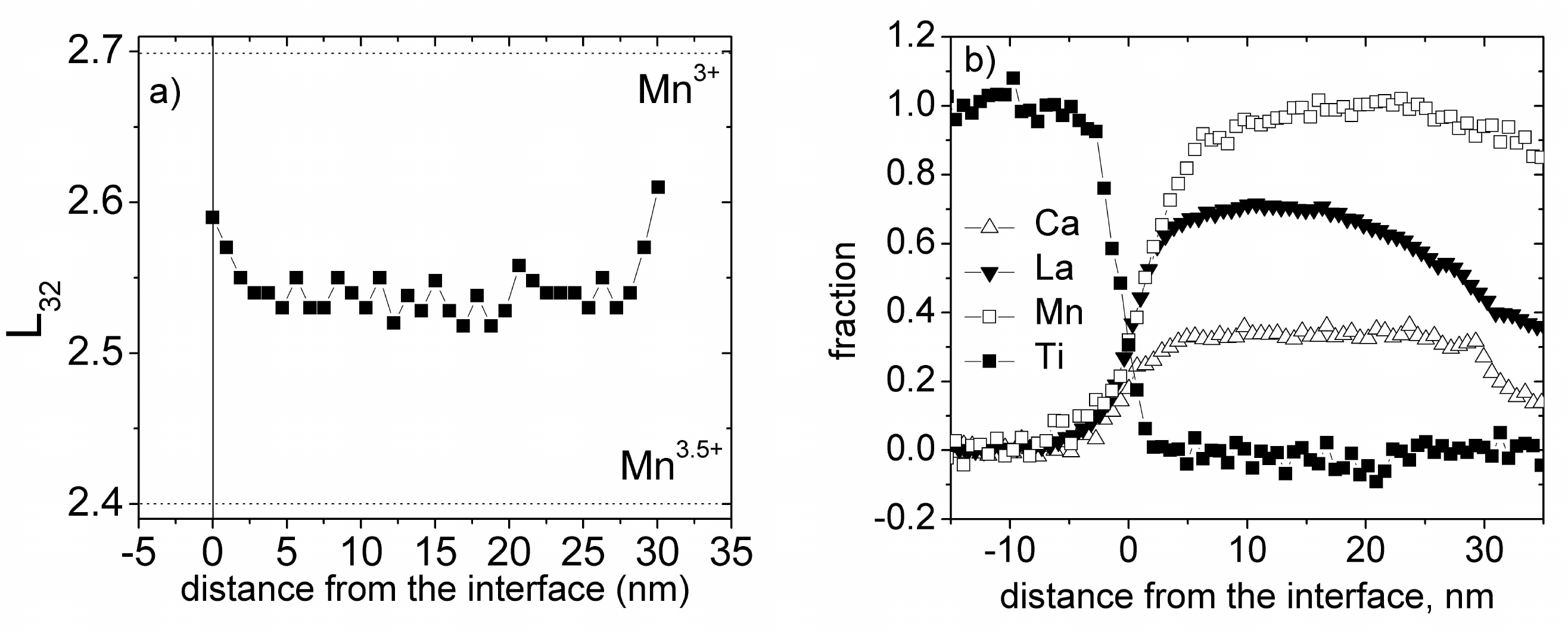}
\centering\caption{a) The calculated L$_{32}$ ratio from the EELS data . Left:
results for sample L(30) on NGO. b) Elemental fraction versus distance from the
interface for the same sample. Plotted are Ca (open triangles), La (filled
triangles), Ti (filled squares), Mn (open squares) }\label{fig6}
\end{figure}

In Fig.\ref{fig4} the calculated L$_{32}$ ratio is shown for several films grown on
STO. Three regions can be discerned. At the film surface (interface with vacuum) the
Mn-valence is reduced towards 3+. In the bulk of the film, particularly visible in
L(47), the L$_{32}$ ratio indicates a Mn-valence close to 3.3+, similar to the value
for bulk LCMO. Close to the interface with the STO substrate a reduction in
Mn-valence is observed. With the probe placed on the first layer the valence is
equal to 3.2+ in both L(6) and L(47). The presence of this reduction is typical for
our LCMO films grown on STO substrates, but the extent of the reduction varies
between 2 - 5 nm for different films. A similar reduction is also found for
L(7)$_{mis}$, grown on 1$^{\circ}$ miscut STO. The step edges on the STO surface
appear to have no significant influence on the Mn-oxidation state near the substrate
interface. Therefore, these properties are general features of LCMO films grown by
sputtering on STO substrates. Fig.\ref{fig4} also shows sample L(10), however, which
did not show the reduced ratio at the substrate interface. The precise cause of this
remains unclear, especially since HR-TEM does not indicate a different type of
interface. \\
We also used EELS to map the elemental composition across the films. As shown in
Fig. \ref{fig5} for L(6) and L(10), the composition (for the most part) is close to
that of the sputtering target. The apparent interdiffusion of Ti$^{4+}$ into the
film is caused by the limited resolution (determined by the beam size) of the
measurement. Other films showed a similar profile. In particular, we do not find
cation segregation at the interface, as a possible cause for the deviation of the
Mn-valence with respect to the bulk value of the film. One exception was sample
L(47), where we observed La enrichment near the film-substrate interface. However,
since the results for all other samples did not show such segregation, it is likely
that the observed La-enrichment is not correlated with the observed Mn-valence
reduction.
An important question is whether the valence reduction at the interface is caused by
strain. We therefore investigated the Mn-valence profile and elemental composition
in LCMO films grown on NGO substrates as well. The results, plotted in Fig.
\ref{fig6}, show that also for films grown on NGO substrates an increase of the
Mn$^{3+}$/Mn$^{4+}$ ratio is observed at the substrate interface. Also similar to
films on STO we find that there is no cation segregation for the films on NGO
substrates.
We discuss these findings below, but the conclusion from this paragraph is that we
observe a slightly higher (0.1 $e$) Mn valence at the interface, which relaxes to
the bulk value in 2 to 5 nm, and which is not caused by either elemental segregation
or by strain.
\section{Transport properties}\label{transport}
\subsubsection{\textit{Films on flat STO}}
\begin{figure}[t]
\centering\includegraphics[height=8cm, width=9.6cm,angle=0]{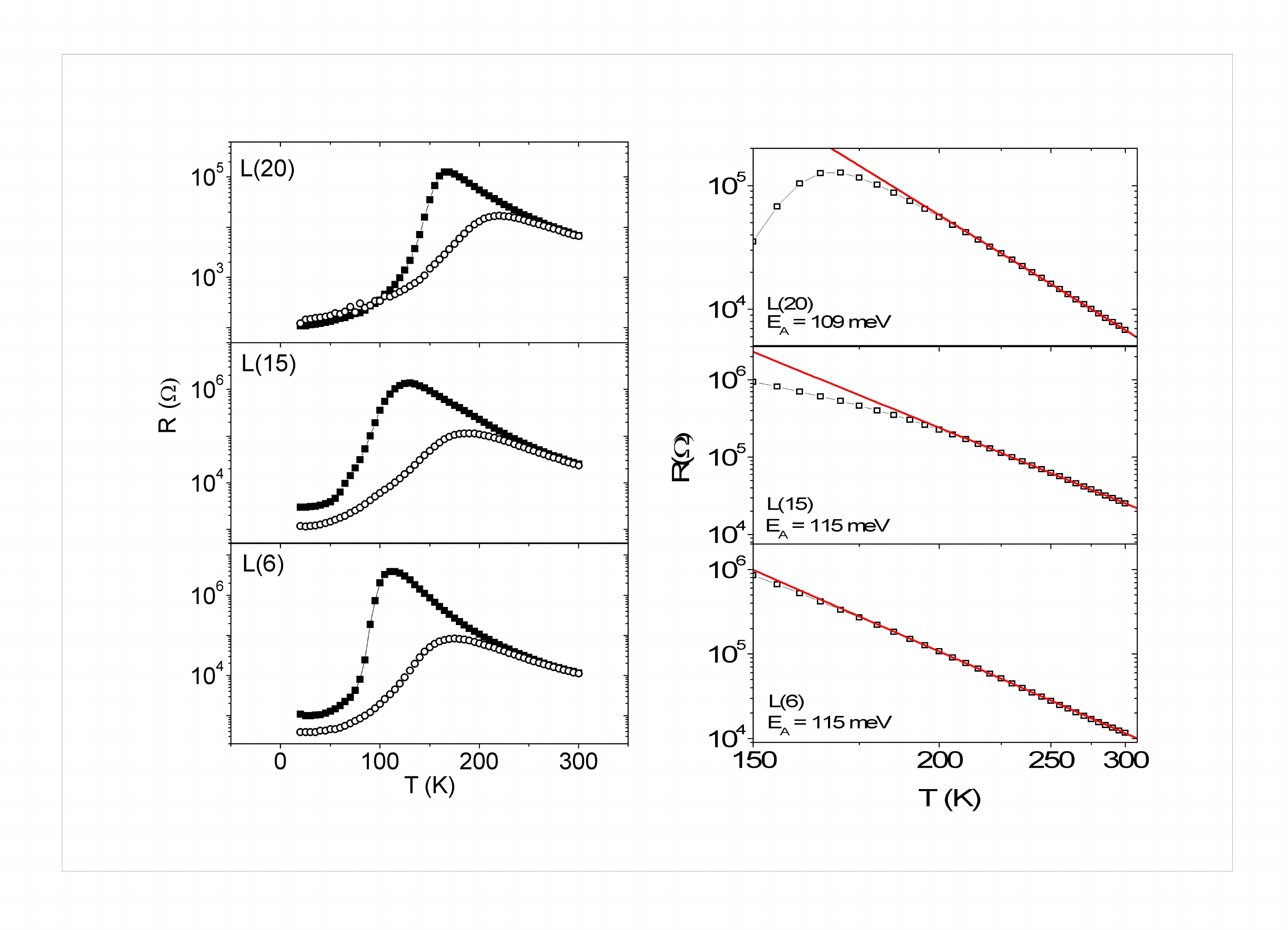}
\centering\caption{(Left) resistance $R$ vs. temperature $T$ for samples L(20) (top
panel), L(15) (center panel) and L(6) (bottom panel) as determined form the I-V
curves at an applied current of 0.1 $\mu$A. Squares: zero field; circles: H$_a$ = 5
T. (Right) log R vs. T in the temperature range 150 - 300 K for the same samples.
Note: the scale on the T-axis is reciprocal in order to show the 1/T-behavior. The
fit to extract the activation energy (E$_A$) of the polaron hopping process is also
shown (solid lines).}\label{fig7}
\end{figure}

\noindent
We measured the $I$-$V$ characteristics in a temperature range of 20 - 300 K for
films with varying thicknesses between 6 - 20 nm, which were linear, except in a
small range around the transition, where weak non-linearities were found. In
Fig.\ref{fig7} we show the temperature dependent resistance R(T) for three LCMO
films grown on flat STO, as determined at an applied current I of 0.1 $\mu$A.
The films show a clear metal-insulator transition accompanied by a resistance drop
of three orders of magnitude. All films show typical CMR effect, a reduction in
resistance of a few orders of magnitude upon application of a 5 T magnetic field.
For L(20) the transition temperature T$_{MI}$, which corresponds to the maximum
resistance value, is 170 K, which occurs approximately 100 K below T$_{MI}$ for bulk
LCMO. T$_{MI}$ is further reduced when the film becomes very thin, L(6) shows the
transition at T$_{MI}$ = 110 K, which is 60 K below T$_{MI}$ for L(20).
\begin{figure}[t]
\centering\includegraphics[width=8.06cm,height=6.4cm,angle=0]{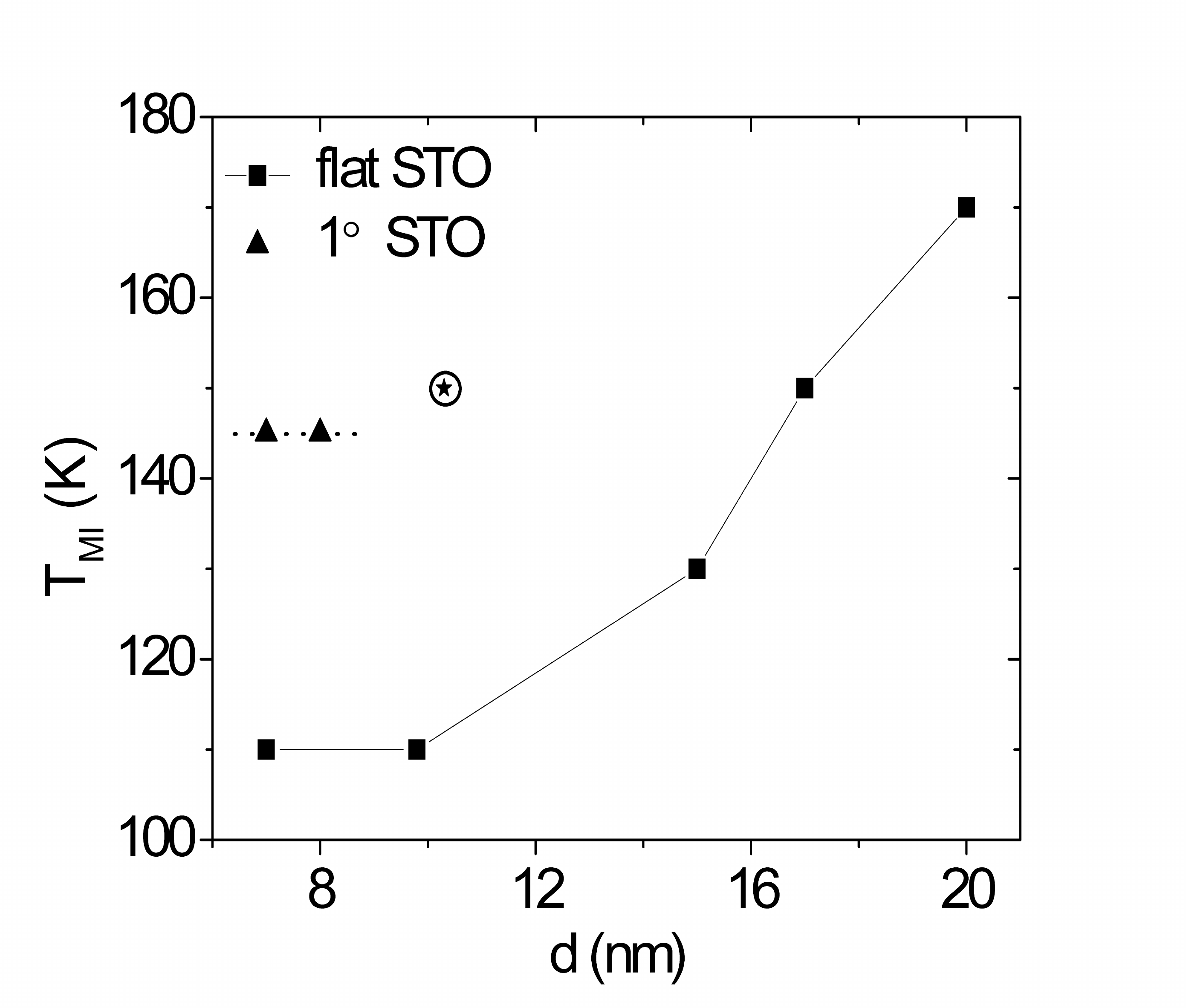}
\centering\caption{The dependence of T$_{MI}$ on film thickness for LCMO films grown
on flat (squares + star) STO and 1$^{\circ}$ misoriented STO (triangles). The star
designates sample L(10). Drawn and dashed lines are guides to the eye. }\label{fig8}
\end{figure}
The dependence on film thickness of T$_{MI}$ is shown in Fig.\ref{fig8}. For films
grown on flat STO the metal-insulator transition is steadily shifted to lower
temperature as the film thickness is reduced. However, the film L(10) (indicated by
$\star$ in Fig.\ref{fig8}) grown on flat STO deviated from this trend. This result
together with the flat Mn valence profile (Fig.\ref{fig4}b implies that L(10) may
have a deviating epitaxial relation with the substrate. There is a connected
observation in the magnetization measurements, which we will discuss beow.

In the paramagnetic state, R(T) is expected to show activated behavior. In
Fig.\ref{fig7}, we plot log R vs. T (with the T-axis reciprocal to show the
1/T-behavior). A linear fit to the data provides the activation energy (E$_A$) for
the polaron hopping process. For most films the high temperature data show activated
behavior with E$_A$ = 110 - 120 meV (see Table \ref{table}), independent of the film
thickness. Deviations typically set in around 1.3x T$_{MI}$ (for example see sample
L(15) center panel Fig.\ref{fig7}). In this respect at least, the thin films do not
behave different from bulk material.

\subsubsection{\textit{Films on 1$^{\circ}$ misoriented
STO}}
\begin{figure}[t]
\centering\includegraphics[width=8.0cm,height=5.66cm,angle=0]{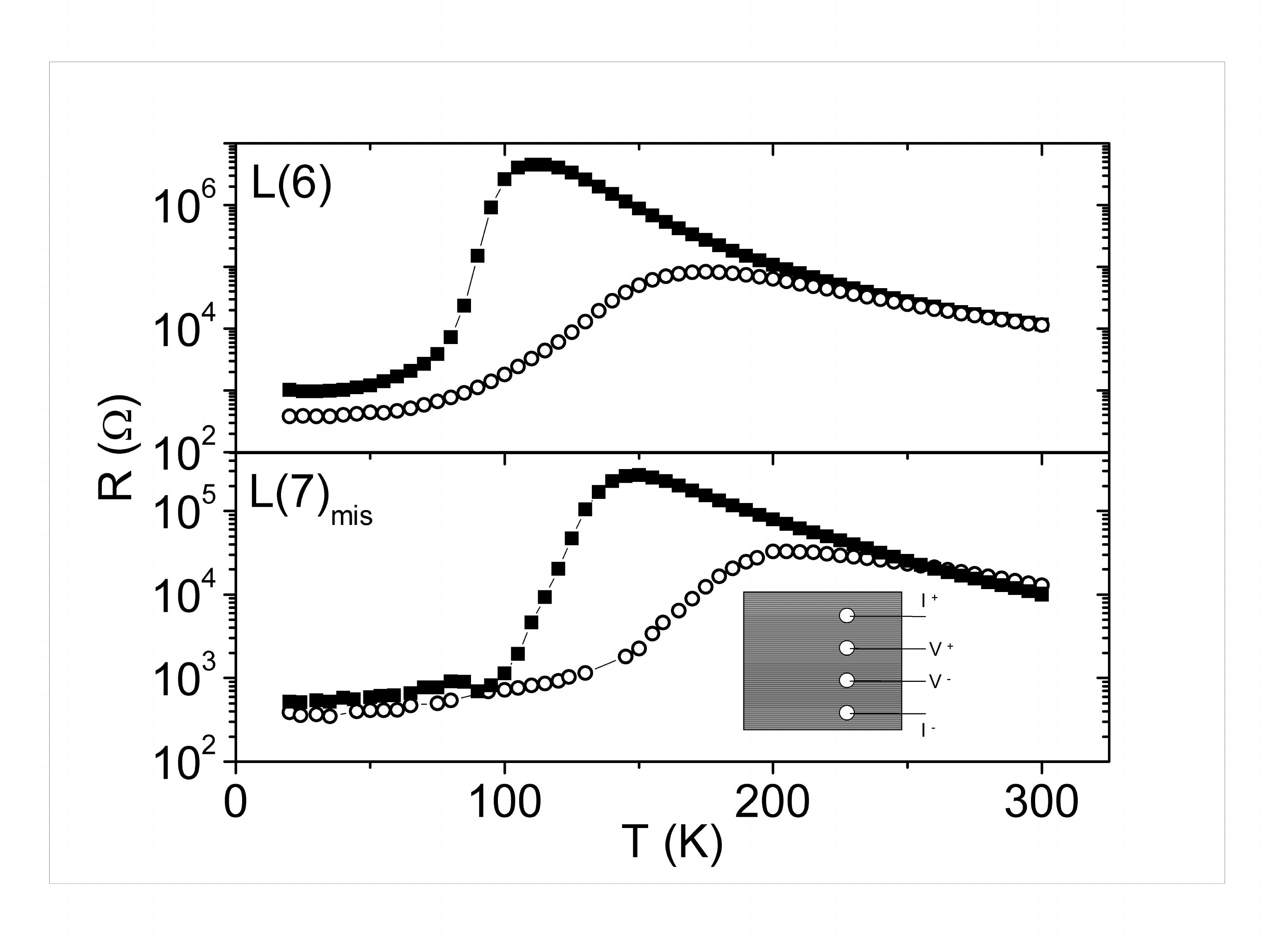}
\centering\caption{Resistance vs. temperature behavior of sample L(6) (also shown in
Fig. \ref{fig7}) and of L(7)$_{mis}$. For the films on misoriented STO the transport
properties were measured with the current directed perpendicular to the step edges
(see inset for measurement geometry). The resistance values were determined form the
I-V curves at an applied current of 0.1 $\mu$A. The squares denote zero field;
circles: H$_a$ = 5 T.}\label{fig9}
\end{figure}

\noindent A novel feature of our studies is the investigation of the effect of
unit-cell high step edges on the STO substrate surface on the thin film properties.
For these films the transport properties were measured in a four-point geometry with
the current directed perpendicular to the step edges (see inset Fig.\ref{fig9}).
Their thickness was determined from HR-TEM micrographs and is 7 and 8 nm
(L(7)$_{mis}$ and L(8)$_{mis}$). R(T) for L(7)$_{mis}$ was measured at $I$ = 0.1
$\mu$A and is shown in Fig. \ref{fig9}, with R(T) of L(6) shown for comparison.
The observed T$_{MI}$ for both films on 1$^{\circ}$ STO is 145 K, so T$_{MI}$ is
much less reduced than for films on flat STO (see Fig.\ref{fig8}, triangles).
However, from the HR-TEM and RSM characterization it is clear that the films on
misoriented STO are fully epitaxial across the entire film thickness. The lack of
reduction in T$_{MI}$ for films on misoriented STO is step-induced but not due to
the loss of epitaxial relation with the substrate. It has been shown before that
strain relaxation in these materials may occur in the form of dislocations in the
film \cite{lippmaa4}. From HR-TEM we did not observe any dislocations in our thin
films, however, point defects should still be present and the amount is possibly
enhanced by the presence of the steps.

\section{Magnetic properties}\label{mag4}
\begin{figure}[t]
\begin{tabular} {cc}
\includegraphics[width=4cm]{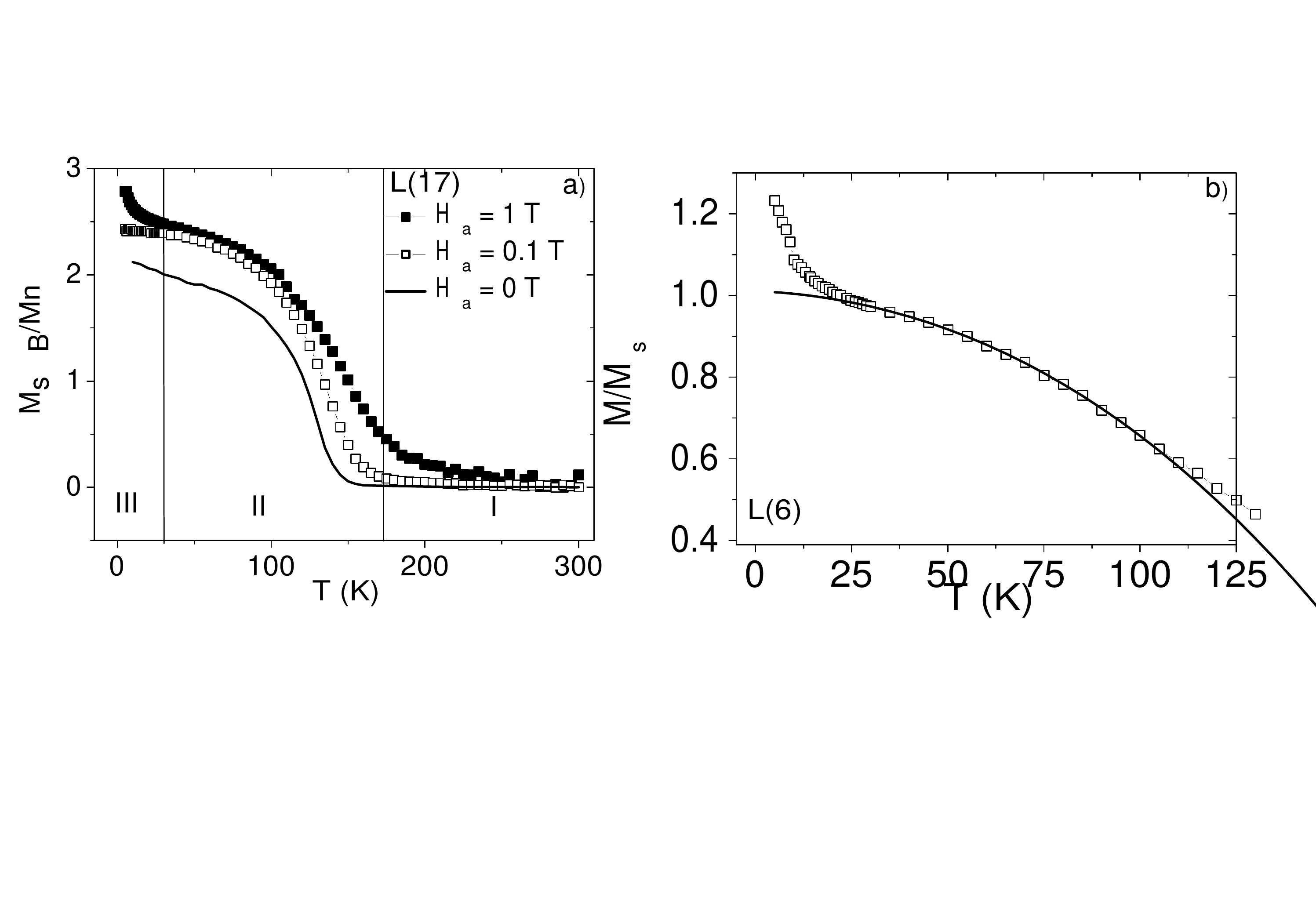} &
\includegraphics[width=4.5cm]{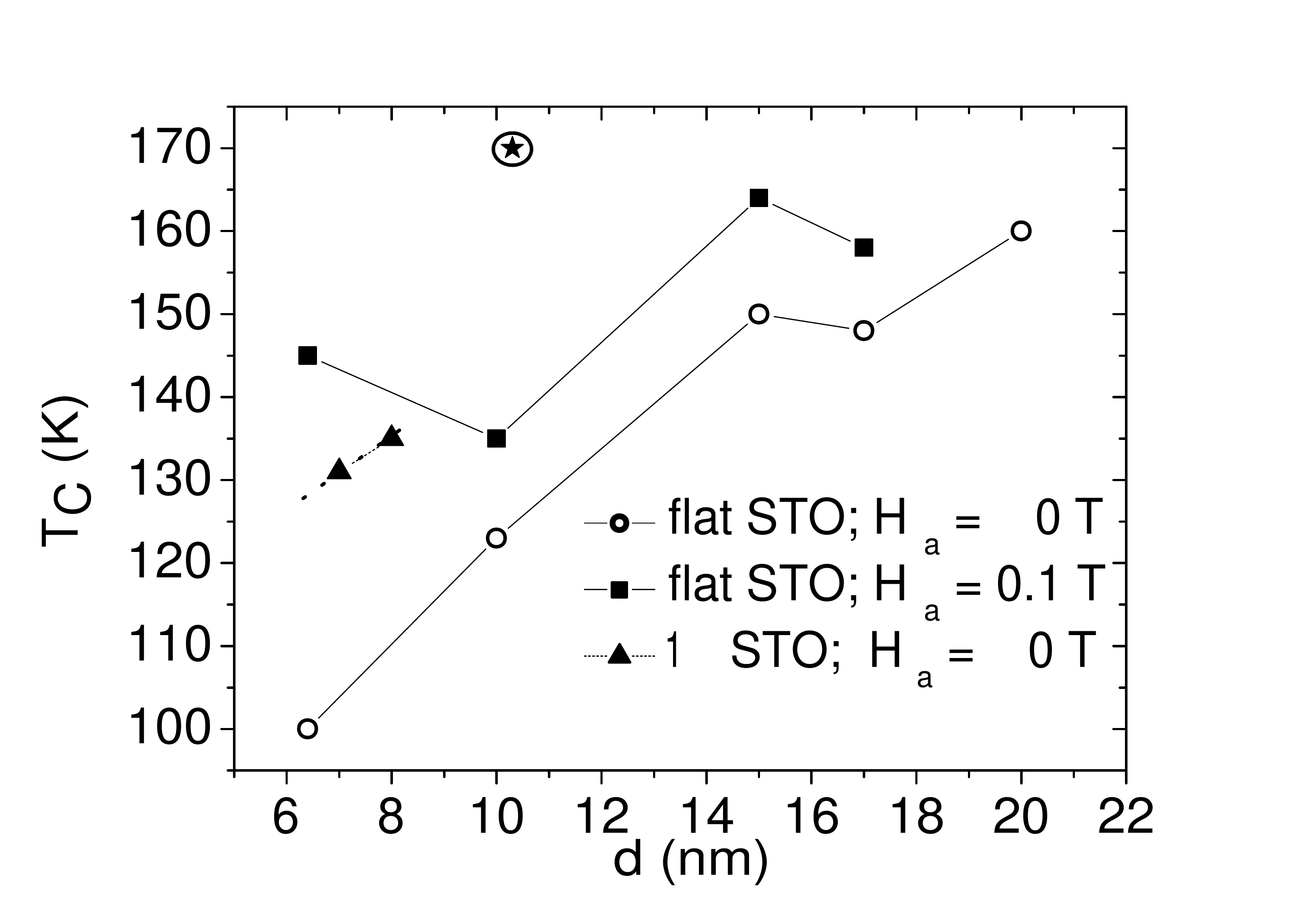} \\
\end{tabular}
\centering\caption{a) M vs. T behavior for L(17). The magnetization was measured in
H$_a$ = 1 T (closed squares), H$_a$ = 0.1 T (open squares) and H$_a$ = 0 T (line).
b) The Curie temperature as function of film thickness, determined from M vs T at
H$_a$ = 0 T (circles) and H$_a$ = 0.1 T (squares). The triangles show T$_C$ for two
films grown on 1$^{\circ}$ STO and the star shows T$_C$ for sample L(10). The drawn
and dashed lines are guides to the eye. }\label{fig10}
\end{figure}

\begin{figure}[b]
\begin{tabular}{cc}
\includegraphics[width=4cm,]{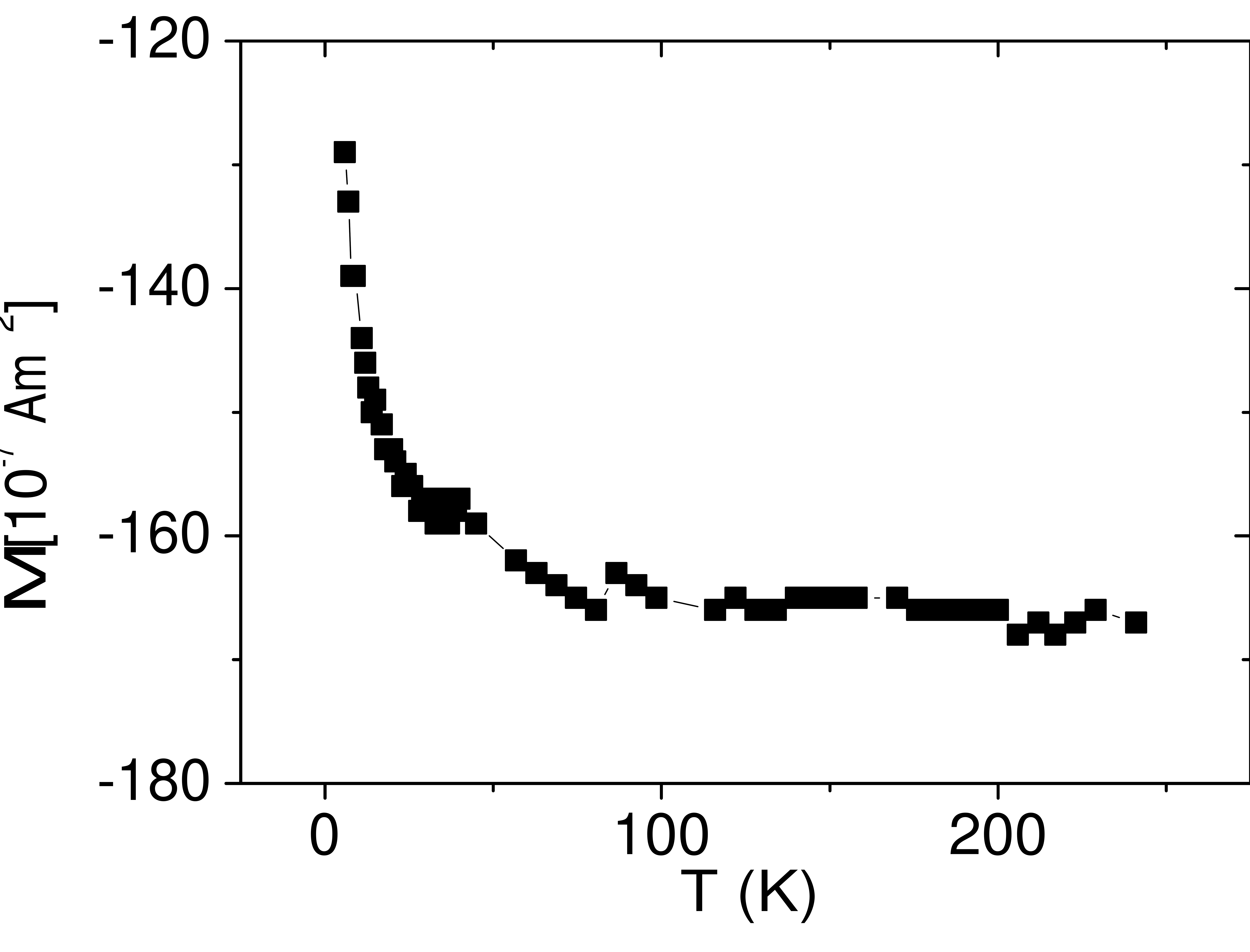} &
\includegraphics[width=4cm]{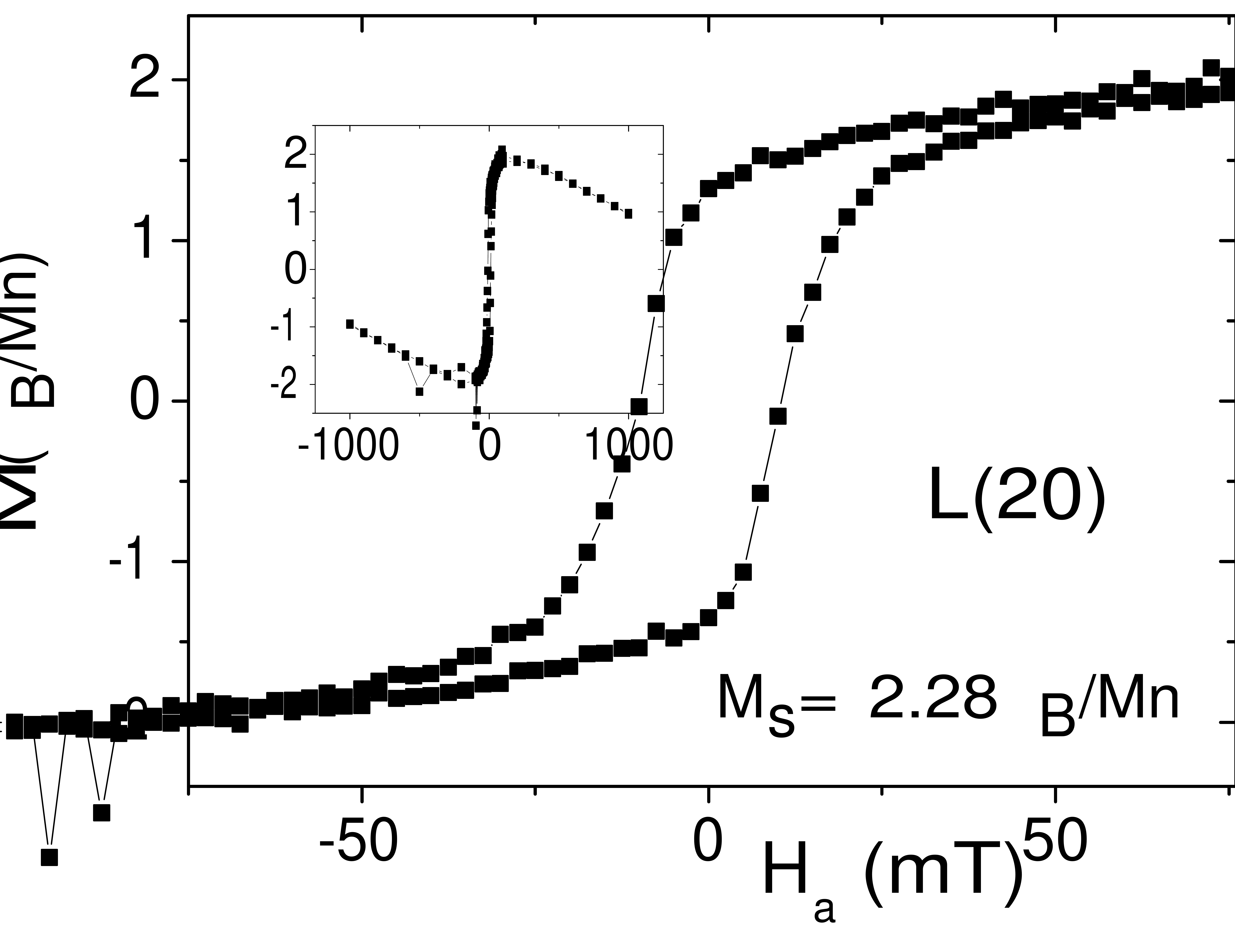} \\
\end{tabular}
\centering\caption{Left: Magnetization as function of temperature for a bare STO
substrate measured in H$_a$ = 1 T. Right: Magnetization as function of applied
magnetic field H$_a$ for the 20~nm thick LCMO film on flat STO. The saturation
magnetization $M_s$ at 1~T is 2.28~$\mu_B$ per Mn-ion. The inset shows M(H$_a$)
measured up to 1~T.}\label{fig11}
\end{figure}

\noindent Here we present the
magnetization behavior of the as-grown films on flat and misoriented STO substrates.
We have not measured the magnetic properties of the films on NGO because the
substrate gives a large paramagnetic background due to the presence of the magnetic
Nd$^{3+}$-ion. Typical behavior of the magnetization $M$ vs. $T$ measured in
magnetic fields of H$_a$ = 0, 0.1~T, 1~T is shown in Fig.\ref{fig10}a for L(17). The
Curie temperature was determined from M vs. T, measured in zero magnetic field and
at H$_a$ = 0.1 T, by taking the intercept of the constant high temperature
magnetization with the linearly increasing M(T).
When the film thickness is reduced, we observe that T$_C$ is also shifted to lower
temperature as shown in Fig. \ref{fig10}b, but continues to coincide with the
temperature of the metal-insulator transition. However, when T$_C$ is extracted from
the measurements in H$_a$ = 0.1 T the Curie temperature is found slightly higher,
and aapears to flatten of at low thickness.  Probably when measured in still higher
fields (for example, H$_a$ = 0.3 T as in Ref.\cite{aarts98}) T$_C$ would be
independent of film thickness. For the measurement in H$_a$ = 1 T we assume that the
magnetization is saturated. In the M(T) behavior at 1~T (see Fig. \ref{fig10}a) the
magnetization shows a sudden increase below T = 30 K. The relative strength of this
upturn increases as the film thickness is reduced but the temperature at which the
upturn starts is constant. This feature is not an intrinsic feature of the LCMO thin
films. From Fig.\ref{fig11} it becomes clear that the $M$ vs. $T$ of a bare STO
substrate also shows an upturn below T = 30 K. Apparently, at low T a paramagnetic
contribution ($\chi$ = C/T, with C the Curie constant) dominates but disappears into
the diamagnetic background above T = 30 K. We surmise that the emergent
paramagnetism is due to the presence of impurities and/or defects in the bulk of the
substrate.
We have also investigated the field dependent magnetization at T = 10 K, shown for
L(20) in Fig.\ref{fig11}b. From these results we can extract the coercive field and
the value for the saturation magnetization (M$_s$) as function of film thickness.
The coercive field varies between +/-6 and +/- 15 mT for the different samples. The
magnetization is given in units of $\mu$$_B$ per Mn-ion and can be described, at
high fields, as M = M$_s$ + $\chi$H (see inset Fig.\ref{fig11}b). The plotted data
in the main graph is corrected for the diamagnetic and paramagnetic contribution of
the STO substrate by subtracting $\chi \times H_a$ determined at high fields. The
values for M$_s$ are determined after correction by taking the value for M at H$_a$
= 1 T. The theoretical saturation magnetization for bulk LCMO would be 3.7 $\mu$$_B$
per Mn-ion. For our films we observe slightly fluctuating values for M$_s$ but
always low compared to the expected bulk value. For L(20) in Fig.\ref{fig11} the
value is 2.28~$\mu_B$/Mn-ion. The lower values indicate that a magnetically dead or
weak layer is generally present in our LCMO thin films.

\section{Discussion}\noindent
The first point to discuss is the reduced Mn-valence we find near the film-substrate
interface (also observed in La$_{0.7}$Sr$_{0.3}$MnO$_3$/STO \cite{riedl3}), which is
not caused by oxygen deficiency. Since the film is grown at high temperature,
diffusion of Ti$^{4+}$ into the film could occur. The presence of the Ti$^{4+}$
introduces a higher charge and as a result the Mn$^{3+}$/Mn$^{4+}$ ratio has to
increase to compensate. Characterization of the elemental composition (see
Fig.\ref{fig5}) and experiments reported by others \cite{samet3} show that
interdiffusion does not occur beyond a few atomic planes. The EELS analysis of the
elemental composition show that in general the films have a composition close to
that of the sputtering target, making it likely that the deviating Mn-valence is not
associated with cation segregation. Since films grown on NGO show similar Mn-valence
profiles the reduction appears to depend on the epitaxial relation between film and
substrate but not on epitaxial strain. We suggest that the Mn-valence at the
interface is affected by the termination of the substrate. In the case of STO
substrates the termination can be Sr$^{2+}$O$^{2-}$ or Ti$^{4+}$O$_2^{4-}$, both
terminations are neutral. The manganite growth can start with either a
[(La$_{0.7}$Ca$_{0.3}$)O]$^{0.7+}$ layer (LaO has charge 1+ and CaO is neutral) or a
MnO$_2$ layer which can have charge 1- or be neutral depending on the Mn-valence
state. A charge mismatch could therefore arise (Madelung potential) across the
LCMO-STO interface.
\begin{table}[t]
\begin{tabular}{|c|c|c|c|c|c|c|c|c|}\hline
  Sample&\textit{d}& \textit{a$_{in}$} & \textit{a$_{out}$} & T$_{MI}$&T$_C$&  E$_A$& M$_s$& d$_{dead}$ \\
     &nm& ($\dot{A}$) & ($\dot{A}$) & (K) & (K) &(meV) &   ($\mu$$_B$/Mn)& nm \\ \hline
   L(20) &20 &3.90&3.8113 & 170& 160 &109&   2.28& 7.6 \\
   L(17) &17 &3.91&3.8135 & 150& 148 &114&   2.89 &3.7\\
   L(15) &15 &3.90&3.8097&   130&150 &115&  1.96 &7.0\\
   L(10) &10.3& 3.90&3.796 & 150& 170 & 114& 3.50 &0.5\\
   L(6) &6.4& 3.91& 3.7979& 110& 100 &115 &  3.30 &0.7\\
   L(8)$_{mis}$ &8 &-& -&  145 &135 & 119 &  2.41 &2.8\\
   L(7)$_{mis}$ &7 &3.90& 3.7911 &  145 &131 & 116 & 2.22 &2.8\\
   L(30)$_{NGO}$ &30& -& -&  285 &- & - &  -&-\\
   L(10)$_{NGO}$&10& -& -&  220 & - & - &  - &-\\ \hline
\end{tabular}
\centering\caption{Summary of the measured values for the thin films presented in
this paper. Samples L(20)-L(6): on flat STO; L(8)$_{mis}$, L(7)$_{mis}$: on
1$^{\circ}$ miscut STO; L(30)$_{NGO}$, L(10)$_{NGO}$: on NGO. The measured variables
are, \textit{d}: film thickness, \textit{a$_{in,out}$}: in-plane, out-of-plane
lattice parameters, T$_{MI}$: MI transition temperature, T$_C$: Curie temperature,
E$_A$: polaron hopping activation energy, M$_s$: saturation magnetization at T = 10
K at 1~T ($\mu_B$/Mn), the corresponding dead layer thickness
d$_{dead}$.}\label{table}
\end{table}
The reduction of the Mn-valence at the interface shows that growth starts with the
[(La$_{0.7}$Ca$_{0.3}$)O]$^{0.7+}$ layer and that the interface polarization leads
to a charge compensation layer with varying thickness (2 - 5 nm); the variation
could be caused by different TiO$_2$/SrO termination ratios of the substrate
surfaces. This also explains the absence of Mn-valence deviations for films on (110)
STO, as reported by Estrad\'{e} et al \cite{fontcuberta3}. The termination layer of
(110) STO is either Sr$^{2+}$Ti$^{4+}$O$^{2-}$ (total charge: 4+) or O$_2^{4-}$. The
first layer of the film would be either La$^{3+}$Mn$^{3+}$O$^{2-}$ or
Ca$^{2+}$Mn$^{4+}$O$^{2-}$, which both have a total charge of 4+, or O$_2^{4-}$. The
result is that LCMO (110) grown on STO (110) would not have a charge mismatch
between film and substrate and therefore does not show a Mn-valence reduction at the
substrate interface. The similar Mn-valence profiles in films on NGO substrates are
explained in the same way. We use NGO (100) (pseudocubic notation \cite{ngo4}) which
has either Nd$^{3+}$O$^{2-}$ or Ga$^{3+}$O$_2^{4-}$ termination layers. The
LCMO(001) film will start the growth with a layer with charge $\pm$(1-x) (x is the
Ca-doping) as described above. The resulting Madelung potential again leads to the
observed Mn-valence profiles. The only sample for which the Mn-valence reduction is
absent (sample L(10)) possibly shows a deviating film orientation w.r.t. the STO
substrate. It is not a priori clear why there is no charge mismatch for this sample
since the STO would be neutral and the first film layer should still have a charge.
One could imagine that the growth started with a CaMnO$_3$ layer for which the
charge mismatch with the substrate would indeed be equal to 0. The resolution of the
EELS measurement to determine elemental composition would be insufficient to detect
such a small Ca-segregation at the STO interface. We were unable to further
investigate this effect since the absence of the compensation layer was only found
once and is not a general feature of the LCMO thin films. Furthermore, we have
observed that the presence of unit-cell high steps on the STO surface do not
significantly influence the microstructure, the Mn-valence profile or the strain
state of the film.
\\
Next we couple this finding an interface induced Mn valence profile to the other
measured properties, summarized in Table~\ref{table}. For films on flat STO,
T$_{MI}$ and T$_{C}$ coincide, decrease with decreasing thickness to a final value
around 110~K at a thickness below 10~nm. This agrees very nicely with the data
presented by Bibes {\it et al.} \cite{bibes01} on similar strained films and
confirms once more the effect of strain on T$_{MI}$. It is tempting to also connect
our finding of a valence change to their conclusion from NMR data, that a
non-ferromagnetic insulating (NFI) phase exists, at least partly, in a region up to
5~nm from the interface. On the other hand, our thinnest films show an almost full
saturation magnetization in 1~T. This suggest that antiferromagnetic interactions
are significant in the valence-changing layer, but that they are still easily
brought to saturation.
\\
The films with thicknesses of 15~nm and 20~nm show a much lower saturation
magnetization (even in 1~T) than might be expected. Since the valence-changing layer
does not seem to play a role here, this appears to be due to growth-induced defects,
again antiferromagnetic in nature, but now not easily saturated. This is a similar
conclusion as was reached by us before (\cite{aarts98}, but in contrast to e.g.
Ref.\cite{bibes01}, and appears to depend on the exact growth conditions of the
films. Interestingly, the films on miscut substrates, where relaxation has set in
more strongly, also show low values for $M_s$. This reinforces the argument that the
step edges lead to disorder in the in the form of local defects \cite{lippmaa4}),
which then also enhances T$_{MI}$.

\section{Conclusion}\noindent
In this paper we have shown that a layer with lower Mn valence is present close to
the interface between a film of La$_{0.7}$Ca$_{0.3}$MnO$_3$ and a SrTiO$_3$ or
NdGaO$_3$ substrate. It presence is not related to strain but, as we argue, the
result of charge compensation. Magnetically we find reduced values for the
saturation magnetization for the strained films, but the apparent lack of magnetic
moment cannot be exclusively ascribed to the presence of the charge compensation
layer. The introduction of step edges onto the SrTiO$_3$ surface leads to a
metal-insulator transition temperature T$_{MI}$ which is less reduced than expected
from the thickness of the film, which we argue to be due to additional disorder in
the form of local defects.

\section{Acknowledgements}
We thank I. Komissarov for discussions, and J.A. Boschker (University of Twente) for
performing the RSM measurements. This work was part of the research program of the
Stichting voor Fundamenteel Onderzoek der Materie (FOM), which is financially
supported by NWO.


\begin{thebibliography}{99}
%
\bibitem{tokura06} Y. Tokura, Rep. Prog. Phys. {\bf 69} 797 (2006); see also Y.
Tokura, 'Colossal Magnetoresistive Oxides', Gordon and Breach Science Publishers
(2000).
%
\bibitem{dagotto08} E. Dagotto, S. Yunoki, C. Sen, G. Alvarez and A. Moreo,
J. Phys.: Condens. Matter {bf 20} 434224 (2008); see also E. Dagotto, 'Nanoscale
phase separation and Colossal Mgnetoresistance', Springer Series in Solid-State
Sciences {\bf 136} (2003).
%
\bibitem{salamon01} M. B. Salamon, Rev. Mod. Phys. {\bf 73}, 583 (2001).
%
\bibitem{lynn07} J.W. Lynn, D. N. Argyriou, Y. Ren, Y. Chen, Y. M. Mukovskii, and
D. A. Shulyatev, Phys. Rev. B \textbf{76}, 014437 (2007).
%
\bibitem{aarts98} J. Aarts, S. Freisem, and R. Hendrikx, and H.W.
Zandbergen, Appl. Phys. Lett. \textbf{72}, 2975 (1998).
%
\bibitem{doerr06} K. D\"{o}rr, J. Phys. D: Appl. Phys. {\bf 39}, R125–R150 (2006).
%
\bibitem{bibes01} M. Bibes, Ll. Balcells, S. Valencia, J. Fontcuberta, M.
Wojcik, E. Jedryka, and S. Nadolski, Phys. Rev. Lett. {\bf 87}, 067210 (2001).
%
\bibitem{yang04} Z. Q. Yang, R. Hendrikx, J. Aarts, Y. L. Qin and H. W. Zandbergen,
Phys. Rev. B {\bf 70}, 174111 (2004).
%
\bibitem{ohtomo04} A. Ohtomo and H. Y. Hwang, Nature {\bf 427}, 423 (2007).
%
\bibitem{beekman5} C. Beekman, J. Zaanen and J. Aarts Phys. Rev.
Lett. submitted
%
\bibitem{zener4} C. Zener, Phys. Rev. \textbf{82}, 403 (1951)
%
\bibitem{machida4} A. Machida, M. Itoh, Y. Moritomo, S. Mori, N. Yamamoto ,K.
Ohoyama,and A. Nakamura, Physica B \textbf{281 $\&$ 282}, 524
(2000)
%
\bibitem{tokura4} Y. Tokura and Y.Tomioka, C, J Mag. Mag. Mat. \textbf{200}, 1 (1999)
%
\bibitem{pressure4} H.Y. Hwang, T.T.M. Palstra, S.-W. Cheong, and
B. Batlogg, Phys. Rev. B, \textbf{52}, 15046 (1995) \textbf{39}, R125 (2006)
%
\bibitem{lat} The pseudocubic lattice parameters $a$ (in-plane) and $b$ (out-of-plane) as determined from
Reciprocal Space Mapping are shown in Table \ref{table}.
%
\bibitem{yang3a} Z.Q. Yang, R. Hendrikx, J. Aarts, Y.L. Qin and
H.W. Zandbergen, Phys. Rev. B, \textbf{70}, 174111 (2004)
%
\bibitem{thickness3} Specimen thickness should not be confused with the film thickness.
%
\bibitem{Gianluigi3} Gianluigi et al., J. Microscopy, \textbf{180}, 211 (1995)
%
\bibitem{lippmaa4} K. Terai, M. Lippmaa, P. Ahmet and T. Chikyow, T. Fujii, H. Koinuma, and M. Kawasaki, Appl. Phys. Lett.
\textbf{80}, 4437 (2002)
%
\bibitem{riedl3} T. Riedl, T. Gemmin, K. D$\ddot{o}$rr, M. Luysberg, and K. Wetzig, Microsc.
Microanal. \textbf{15}, 213 (2009)
%
\bibitem{samet3} L. Samet, D. Imhoff, J.-L. Maurice, J.-P. Contour,
A. Gloter, T. Manoubi, A. Fert, and C. Colliex, Eur. Phys. J. B. \textbf{94}, 179
(2003)
%
\bibitem{fontcuberta3} S. Estrad\'{e}, J. Arbiol, F.Peir\'{o}, I.C.
Infante, F. S\'{a}nchez, J. Fontcuberta, F. de la Pe\~{n}a, M.
Walls, and C. Colliex, Appl. Phys. Lett. \textbf{93}, 112505
(2008)
%
\bibitem{ngo4} Note that different orientations of the NGO
substrate lead to different strain states in the film. Here we use
NGO (100), cubic notation, which is (110) in orthorhombic
notation.
%

\end{thebibliography}
\end{document}